\documentclass[11pt]{article}

\usepackage[final]{acl}

\usepackage{times}
\usepackage{latexsym}

\usepackage[T1]{fontenc}

\usepackage[utf8]{inputenc}

\usepackage{microtype}

\usepackage{inconsolata}

\usepackage{hyperref}
\usepackage{url}
\usepackage{wrapfig}
\usepackage{subcaption}
\usepackage{graphicx}
\usepackage{arydshln}
\usepackage{booktabs} 
\usepackage{multirow} 
\usepackage{makecell} 
\usepackage{tabularx}
\usepackage{enumitem}
\usepackage{array}  
\usepackage{adjustbox}
\usepackage{tcolorbox}
\usepackage{xcolor}
\usepackage{colortbl}  
\usepackage{wrapfig}
\usepackage{tcolorbox}

\definecolor{red}{rgb}{0.99, 0.02, 0.02}
\definecolor{mgreen}{RGB}{0,200,0}

\title{\textsc{SweRank+}: Multilingual, Multi-Turn Code Ranking\\for Software Issue Localization}

\author{Revanth Gangi Reddy\thanks{Equal Contribution. Work done during Revanth's internship at Salesforce AI Research.}\hspace{0.1em}$^{1,2}$\hspace{0.5em}Ye Liu\footnotemark[1]\hspace{0.1em}$^2$\hspace{0.5em}Wenting Zhao\hspace{0.1em}$^2$\hspace{0.5em}JaeHyeok Doo$^3$\hspace{0.5em}Tarun Suresh$^1$\\\textbf{Daniel Lee}\hspace{0.1em}$^2$\hspace{0.7em}\textbf{Caiming Xiong}\hspace{0.1em}$^2$\hspace{0.7em}\textbf{Yingbo Zhou}\hspace{0.1em}$^2$\hspace{0.7em}\textbf{Semih Yavuz}\hspace{0.1em}$^2$\hspace{0.7em}\textbf{Shafiq Joty}\hspace{0.1em}$^2$\\
$^1$University of Illinois at Urbana-Champaign\hspace{1em}$^2$Salesforce AI Research\hspace{1em}$^3$KAIST AI\\
  \texttt{revanth3@illinois.edu; \{yeliu,sjoty,syavuz\}@salesforce.com; }}

\begin{document}
\maketitle
\begin{abstract}

Maintaining large-scale, multilingual codebases hinges on accurately localizing issues, which requires mapping natural-language error descriptions to the relevant functions that need to be modified. However, existing ranking approaches are often Python-centric and perform a single-pass search over the codebase. This work introduces \textsc{SweRank+}\footnote{Code and models will be released here: \url{https://github.com/SalesforceAIResearch/SweRank}}, a framework that couples \textsc{SweRankMulti}, a cross-lingual code ranking tool, with \textsc{SweRankAgent}, an agentic search setup, for iterative, multi-turn reasoning over the code repository. \textsc{SweRankMulti} comprises a code embedding retriever and a listwise LLM reranker, and is trained using a carefully curated large-scale issue localization dataset spanning multiple popular programming languages. \textsc{SweRankAgent} adopts an agentic search loop that moves beyond single-shot localization with a memory buffer to reason and accumulate relevant localization candidates over multiple turns. Our experiments on issue localization benchmarks spanning various languages demonstrate new state-of-the-art performance with \textsc{SweRankMulti}, while \textsc{SweRankAgent} further improves localization over single-pass ranking. 
\end{abstract}

\section{Introduction}
\label{sec:intro}

The maintenance of large-scale software systems constitutes a significant and ever-growing portion of the software development lifecycle. A persistent bottleneck in this process is \textit
{software issue localization}~\cite{wong2016survey}: the task of identifying where in a codebase a fix should be applied for a given bug report or feature request. This task requires mapping natural language descriptions, such as those found in GitHub issues, to specific code elements including files, modules, or functions. As modern code repositories grow in size and complexity to encompass thousands of files across multiple programming languages, manual localization becomes increasingly infeasible. Automating this process can therefore accelerate issue resolution and significantly enhance developer productivity.

\begin{figure}
    \centering
    \includegraphics[width=0.71\linewidth]{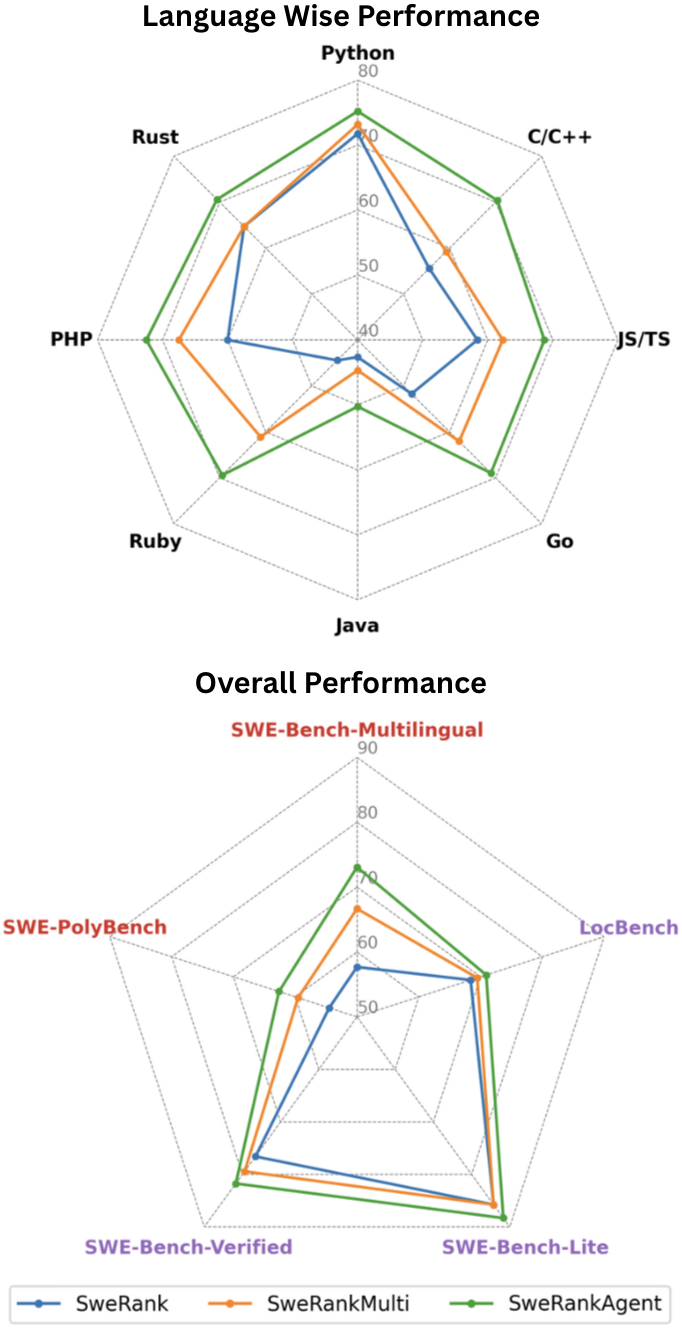}
    \caption{Comparison of function localization accuracy@10 against the \textsc{SweRank} baseline. \textsc{SweRankMulti} shows
significant improvement on multilingual benchmarks (in red) while
maintaining strong performance on Python-specific evaluations (in violet). \textsc{SweRankAgent} further improves over single-pass ranking approaches across the board.}
    \label{fig:final_comparison}
    \vspace{-1em}
\end{figure}

Recent advances in large language models (LLMs) have led to the development of sophisticated \textit{agentic systems}~\cite{yu2025orcaloca, chen2025locagent} capable of navigating complex codebases. These systems operate by issuing commands to read files, search for patterns, and analyze dependencies. While powerful, such approaches often suffer from high latency and substantial computational overhead. To improve efficiency, \textsc{SweRank}~\cite{reddy2025swerank} reformulates issue localization as a \textit{retrieve-and-rerank} problem, leveraging specialized bi-encoder retrievers and LLM rerankers trained on curated datasets~\cite{suresh2024cornstack}. This approach has achieved high precision while substantially reducing computational cost.

Despite its success, \textsc{SweRank} faces two key limitations that reflect the evolving needs of modern software engineering. First, its scope is largely confined to Python-dominated repositories, whereas real-world enterprise systems are inherently \textit{multilingual}, comprising interconnected components written in diverse programming languages. A practical localization tool must therefore generalize across languages. Second, \textsc{SweRank}'s single-pass ranking design may be inadequate for complex issues that demand \textit{iterative reasoning} or involve changes dispersed across multiple, loosely coupled functions. To overcome these challenges, we introduce \textsc{SweRank+}, a comprehensive framework for multilingual, multi-turn issue localization. 

Our first component, \textsc{SweRankMulti}, extends \textsc{SweRank} to a multilingual setting. It includes a suite of retriever and reranker models trained on \textsc{SweLocMulti}, a newly curated, large-scale dataset containing high-quality issue–code pairs spanning multiple popular programming languages. \textsc{SweRankMulti} enables high-accuracy localization across heterogeneous repositories, marking the first such effort in this domain.

Next, we present \textsc{SweRankAgent}, a novel agentic framework that moves beyond single-shot localization.  Instead of relying on a single retrieval pass, the agent adopts an iterative process: it begins with an initial hypothesis and progressively refines its understanding through multiple turns. Each turn allows it to gather new evidence, narrow the search space, and focus on the most relevant code regions. This multi-turn reasoning process mirrors how human developers investigate issues--starting with a broad, symptom-based exploration and converging toward the root cause. Specifically, \textsc{SweRankAgent} is motivated by three key observations: (A) \textbf{Initial searches can be misleading}--the most obvious function in a traceback is often not the true source of the bug; (B) \textbf{Context is built over turns}--early retrievals reveal important structural clues about the codebase; (C) \textbf{Iterative refinement leads to precision}--repeated reasoning helps transition from symptom-level to cause-level localization.

\textsc{SweRankAgent} integrates \textsc{SweRankMulti} as a specialized retrieval tool within this iterative process.  It can issue multiple search queries, maintain a memory buffer of intermediate results, and reason over aggregated evidence to produce more accurate localizations.  This hybrid design combines the efficiency of high-quality retrievers with the depth of agentic reasoning, enabling it to solve complex localization problems that remain intractable under single-pass methods.

Through extensive experiments, we show that \textsc{SweRankMulti}  establishes new state-of-the-art performance for issue localization across multiple programming languages~\cite{Rashid2025SWEPolyBenchAM, Zan2025MultiSWEbenchAM, yang2025swesmith}, while maintaining competitive results on Python benchmarks~\cite{jimenez2023swe, chen2025locagent}. Furthermore, the iterative reasoning in \textsc{SweRankAgent} consistently outperforms single-pass ranking on issue localization tasks. Figure~\ref{fig:final_comparison} shows our experimental results. In summary, our contributions are:
\begin{itemize}
    \item We present the first framework to address issue localization in a multilingual setting. 
    \item  We introduce \textsc{SweRankMulti}, trained on 10 programming languages, achieving state-of-the-art multilingual code ranking performance for issue localization.  
    \item We propose \textsc{SweRankAgent}, an iterative, multi-turn localization framework that further improves over single-pass ranking.  
\end{itemize}

\section{Related Work}

\label{sec:related_work}

\subsection{Software Issue Localization} Software issue localization (fault localization) aims to identify the specific code regions responsible for software defects. Traditional approaches include spectrum-based methods~\citep{jones2005empirical} and state-based methods~\citep{zeller2002isolating}, which analyze program executions and state differences to isolate potential fault sites. Learning-based methods such as DeepFL~\citep{li2019deepfl} extended these ideas by integrating multiple fault signals using deep representations. Recent approaches to fault localization~\cite{torun2025past, chang2025bridging, yaraghi2025black} have leveraged LLM-based techniques to improve the precision and efficiency of identifying faulty program elements by incorporating semantic reasoning and retrieval-augmented inference, moving beyond traditional spectrum- and mutation-based metrics.


The advances in AI for software engineering have spurred the development of LLM-based agentic frameworks designed to perform complex software engineering tasks~\citep{he2025llm,dong2025surveycodegenerationllmbased}. Recent LLM-based agentic approaches tackles issue localization as a planning and searching problem.~\citet{yang2024sweagent} proposed LLM-based framework that leverages multi-turn reasoning and tool invocation for issue resolution and code repair. \citet{chen2025locagent} incorporates graph-guided reasoning for precise function localization, while \citet{yu2025orcaloca} designs a framework specialized for bug localization using search and read actions. 

While these approaches demonstrate the power of agentic architectures, they often involve complex interactions with the codebase, which can incur high latency~\citep{chen2025locagent, yu2025orcaloca}. \textsc{SweRank+} introduces a distinct, hybrid agentic model, \textsc{SweRankAgent}, that combines the efficiency of a specialized retrieval tool with a lightweight, iterative reasoning loop. This enables tackling complex issues that are intractable for single-pass ranking systems while avoiding the high overhead of complex agentic frameworks. 

\subsection{Multilingual Code Understanding}

Recent code LLMs have significantly expanded their code understanding capabilities across programming languages, with models such as Qwen3-Coder~\citep{yang2025qwen3}, and CodeGemma~\citep{codegemmateam2024codegemmaopencodemodels} exhibiting strong zero-shot transfer between languages. However, the integration of multilinguality into software issue localization remains comparatively nascent. Most localization systems remain monolingual, often trained exclusively on Python (e.g., \textsc{SweRank}~\citep{reddy2025swerank}, LocAgent~\citep{chen2025locagent}), limiting their applicability to real-world software ecosystems. \textsc{SweRank}+ bridges this gap as the first localization framework explicitly trained and evaluated on multilingual repositories.


\section{\textsc{SweRankMulti}}
\label{sec:swerankmulti}

We introduce \textsc{SweRankMulti}, a framework for effective software issue localization across diverse programming languages. \textsc{SweRankMulti} adapts the efficient retrieve-and-rerank methodology of \textsc{SweRank} to the multilingual setting, comprising two key components: the \textsc{SweRankEmbedMulti} retriever (\S\ref{sec:SweRankEmbedMulti}), which pre-selects a small set of candidate functions from large codebases most likely relevant to a given issue, and the \textsc{SweRankLLMMulti} reranker (\S\ref{sec:SweRankLLMMulti}), which produces a refined final ranking from the top candidates.

\subsection{\textsc{SweLocMulti} Training Dataset}
\label{sec:swelocmulti}
We create \textsc{SweLocMulti}, a large-scale multilingual dataset curated specifically for issue localization. While the original \textsc{SweLoc} dataset provided a high-quality resource for Python, modern software systems are inherently multilingual. \textsc{SweLocMulti} extends the data collection and filtering pipeline of \textsc{SweLoc} to encompass JavaScript, Java, TypeScript, Ruby, Rust, Go, PHP, C, and C++. Table~\ref{tab:dataset_stats} presents the language-wise distribution of training instances.

\paragraph{Issue Collection:} Following \textsc{SweRank}'s methodology, we identify popular open-source repositories on GitHub for each language, filtering for repositories with at least 40\% code in the target language, over 1,000 stars, and at least one commit in the preceding six months. From this curated set, we extract pull requests (PRs) explicitly linked to GitHub issues that include test file modifications.

\begin{table}[t]
    \centering
    \small
    \vspace{-1.2em}
    \renewcommand{\arraystretch}{1.2}
    \begin{tabular}{l c c c}
    \toprule
    \textbf{Language} & \textbf{\# Repos} & \textbf{\# PRs} & \textbf{\# Instances} \\  
    \midrule
     JavaScript & 104 & 1513 & 4254 \\
     Java & 130 & 5518 & 19239 \\
     TypeScript & 129 & 3882 & 11410\\
     Ruby & 308 & 4244 & 9048 \\
     Rust & 269 & 5879 & 22255 \\
     Go & 114 & 2985 & 11242 \\
     PHP & 206 & 4591 & 16608\\
     C & 74 & 1023 & 4013 \\
     C++ & 278 & 2359 & 7621\\     
    Python & 2448 & 24285 & 49973\\
     \midrule
     Total & 4060 & 56279 & 155663\\
    \bottomrule
    \end{tabular}
    \caption{Distribution of repositories, pull requests (PRs), and training instances across different programming languages in the \textsc{SweLocMulti} dataset.}
    \label{tab:dataset_stats}
\end{table}

\paragraph{Contrastive Data:} Each issue description serves as a query, with modified functions in the corresponding PR treated as positive examples. To ensure the model learns to distinguish fine-grained semantic differences, we employ \textit{consistency filtering} and \textit{hard-negative mining} techniques~\cite{suresh2024cornstack} from \textsc{SweRank}. Specifically, we use a pretrained embedding model (\textsc{SweRankEmbed}-Small) to perform consistency filtering, retaining only examples where the positive function ranks within the top-40 semantically similar functions repository-wide. We then mine hard negatives, which are unmodified functions from the same repository that are semantically similar to the query, creating challenging instances for the contrastive training process.

\subsection{\textsc{SweRankEmbedMulti} Retriever}
\label{sec:SweRankEmbedMulti}
The retriever component, \textsc{SweRankEmbedMulti}, is a bi-encoder model that maps issue descriptions and code functions to dense vector representations in a shared multilingual embedding space. We initialize the model with Qwen3-Embedding~\cite{zhang2025qwen3}, a state-of-the-art text embedding model, and finetune it on \textsc{SweLocMulti} using the InfoNCE contrastive loss~\cite{oord2018representation}, identical to \textsc{SweRankEmbed}'s training objective. 

Given an issue description as the query, the loss function encourages the model to output query embeddings closer to the corresponding positive code function while pushing away from other functions in the training batch, including mined hard negatives. At inference time, the model ranks all repository functions by cosine similarity between their embeddings and the issue embedding, thereby efficiently identifying the most probable fix locations. 

\subsection{\textsc{SweRankLLMMulti} Reranker}
\label{sec:SweRankLLMMulti}

The reranker component, \textsc{SweRankLLMMulti}, refines the initial candidate list from the retriever using listwise reranking. This approach leverages the instruction-following capabilities of large language models and has proven more effective than pointwise~\cite{zhuang2023rankt5,zhuang2024beyond} or pairwise ranking~\cite{qin2024large}.

Training \textsc{SweRankLLMMulti} leverages the novel weakly supervised strategy introduced in \textsc{SweRank}, which enables listwise reranking learning even when only a single positive example is known, without requiring a fully ordered ground-truth list. Each candidate function receives a unique identifier, and the model is trained to generate the identifier of the true positive function as its first token using standard language modeling loss.  This objective effectively aligns model generation with ranking behavior, teaching it to select the most relevant function from the candidate set. Fine-tuning on the multilingual \textsc{SweLocMulti} dataset substantially enhances cross-language generalization and code ranking accuracy, setting a new state of the art for multilingual issue localization, as demonstrated in Section~\ref{sec:swerankmulti_expts}.

\section{\textsc{SweRankAgent}}
\label{sec:swerankagent}

\begin{figure*}[t]
    \centering
    \vspace{-1.1em}
    \includegraphics[width=1.0\linewidth]{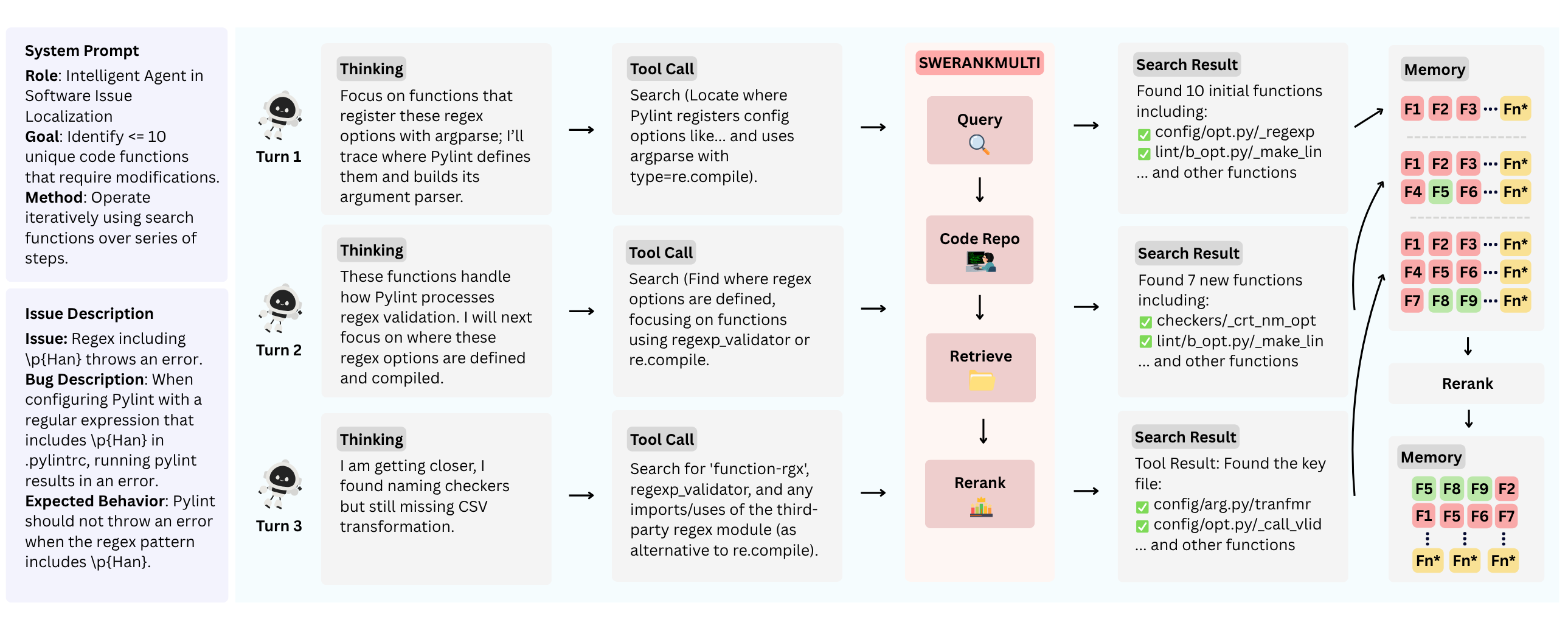}
    \vspace{-2em}
    \caption{\textsc{SweRankAgent} interleaves \textit{Search} actions, which retrieve candidate functions, and \textit{Aggregate} actions, which selectively adds the most relevant functions to a persistent memory. The diagram illustrates a qualitative example of the agent solving a complex issue. The visualization details the agent's multi-turn trajectory (turns 1-3), showing how it iterates through thinking and tool calls to \textsc{SweRankMulti} to progressively narrow down the root cause from a broad search to the target function.}
    \label{fig:swerank_agent} 
    \vspace{-1em} 
\end{figure*}

While \textsc{SweRankMulti} models effectively localize relevant code functions, they operate on a single, static query--the original issue description. This approach performs well for self-contained issues where the initial report provides sufficient signal to identify the fix location. However, many software maintenance tasks involve complex bugs or feature requests whose critical details are not fully captured in the initial description. 
 For example, a bug report may describe only a high-level symptom, and it is only after examining the functions related to that symptom that a developer can pinpoint the downstream component responsible for the fault.

To emulate this human-like, iterative exploratory process, we introduce \textsc{SweRankAgent}, a lightweight and generic agentic framework that extends \textsc{SweRankMulti} with multi-turn search capabilities. Rather than treating issue localization as a single-shot ranking task, \textsc{SweRankAgent} performs iterative searches across multiple turns to progressively accumulate and refine localization evidence. This design enables the agent to decompose complex localization problems into a sequence of smaller, more tractable steps, effectively overcoming the limitations of single-pass retrieval.

\textsc{SweRankAgent} operates as a ReAct-style~\cite{yao2023react} reasoning agent, alternating between reasoning and action-taking. Its iterative operational loop, illustrated in Figure~\ref{fig:swerank_agent}, consists of four key steps: \texttt{Search}, \texttt{Reasoning}, \texttt{Reformulation}, and \texttt{Aggregate}.
\begin{itemize}
    \item \texttt{Search} with \textit{issue description}: The agent's primary interface with the codebase. Given a natural language issue description as the query, this action invokes the \textsc{SweRankMulti} retriever and reranker to obtain the top-$k$ most relevant functions from the code repository.
    \item \texttt{Reasoning} over \textit{retrieved functions}: The agent analyzes the retrieved functions to assess their relevance to the issue and determine whether further exploration is needed. This reasoning step informs subsequent query refinement.
    \item \texttt{Reformulation} for \textit{query update}: After the reasoning step, the agent updates the query based on the previous search and its assessment of the retrieved functions' relevance. This enables the agent to progressively refine its search and improve contextual understanding across iterations.
     \item \texttt{Aggregate} on \textit{candidate functions}: Across turns, the agent accumulates identifiers of relevant functions in an internal memory buffer that serves as a global candidate pool. This enables the agent to capture issues whose fixes span multiple functions that may not co-occur within a single retrieval. Finally, the \textsc{SweRankLLMMulti} reranker is applied to the aggregated pool to produce a globally ranked list of candidate functions.

\end{itemize}

The agent iteratively cycles through \texttt{Search}, \texttt{Reasoning}, and \texttt{Reformulation} until a stopping condition is met--either the maximum number of iterations is reached or no new relevant functions are retrieved. The accumulated candidates are then consolidated and reranked during the final \texttt{Aggregate} step, with the top-10 functions returned as the final localization output. Throughout this process, the underlying LLM guides the agent’s reasoning, enabling it to refine queries, evaluate search results, and dynamically adapt its strategy based on evolving evidence from the codebase.

  \section{Experiments}
\label{sec:expts}

Our experiments are designed to address the following research questions: \textbf{RQ1:} \textit{How effective is training on \textsc{SweLocMulti} compared to the Python-specific \textsc{SweLoc}}? and \textbf{RQ2:} \textit{Can issue localization benefit from multi-turn search}?. To examine \textbf{RQ1}, Section~\S{\ref{sec:swerankmulti_expts}} reports the performance of the \textsc{SweRankMulti} retriever and reranker across multilingual and Python-specific issue localization benchmarks. To investigate \textbf{RQ2}, Section~\S{\ref{sec:swerankagent_expts}} analyzes the effectiveness of multi-turn issue localization using \textsc{SweRankAgent}, and demonstrates its improvements over the single-turn \textsc{SweRankMulti} model. 

\begin{table*}[t]
    \centering
    \renewcommand{\arraystretch}{1.2}
    \small
    \resizebox{1\textwidth}{!}{
    \begin{tabular}{l cc cc cc cc cc}
        \toprule
        
        \multirow{3}{*}{\textbf{Model}} &  \multicolumn{4}{c}{\textbf{Multilingual}} & \multicolumn{6}{c}{\textbf{Python}}\\
        \cmidrule(lr){2-5} \cmidrule(lr){6-11}
        
        &\multicolumn{2}{c}{\textbf{SWE-PolyBench}}  & \multicolumn{2}{c}{\textbf{SWE-Bench-Multilingual}} &\multicolumn{2}{c}{\textbf{SWE-Bench-Lite}} & \multicolumn{2}{c}{\textbf{LocBench}} & \multicolumn{2}{c}{\textbf{SWE-Bench-Verified}}\\
        \cmidrule(lr){2-3} \cmidrule(lr){4-5} \cmidrule(lr){6-7} \cmidrule(lr){8-9} \cmidrule(lr){10-11} 
         & \textbf{Acc@5} & \textbf{Acc@10} & \textbf{Acc@5} & \textbf{Acc@10} & \textbf{Acc@5} & \textbf{Acc@10}& \textbf{Acc@5} & \textbf{Acc@10}& \textbf{Acc@5} & \textbf{Acc@10} \\
        \midrule    
        
        OpenHands (Claude-3.5) &- & - & - & - & 68.25 & 70.07 & - & 59.11 & - & - \\
        LocAgent (Claude-3.5) &- & - & - & - & 73.36 & 77.37 & - & 59.29 & - & - \\
        Gemini-Embedding (unknown) & - & - & 36.75 & 47.44 & 61.31 & 72.26 & 43.04 & 51.43  &  59.74 & 66.96\\
        \midrule
        CodeRankEmbed (137M) & 27.11 & 33.40 & 26.50 & 35.04 & 51.82 & 58.76 & 38.93 & 47.86 & 50.98& 57.99  \\
        Qwen3-Embedding-0.6B & 30.69 & 37.17 & 30.34 & 38.03 & 52.55 & 62.77 & 39.64 & 47.32 & 49.45 & 60.39\\
         \textsc{SweRankEmbed}-Small (137M) & 35.91 & 42.79 & 33.33 & 44.02 & 63.14 & 74.45 & 51.79 & 58.57 &59.74 & 68.49 \\
         \textsc{SweRankEmbedPython}-Small (0.6B) & 37.75 & 43.95 & 36.75 & 47.01 & \textbf{66.79} & 75.18 & 50.00 & 58.04 & 61.93 & 71.12 \\
         \textbf{\textsc{SweRankEmbedMulti}-Small} (0.6B) & \textbf{39.01} & \textbf{47.14} & \textbf{43.16}  & \textbf{52.56} & \textbf{66.79} & \textbf{76.28} & \textbf{51.25} & \textbf{58.93} & \textbf{64.99} & \textbf{71.77} \\
         \midrule
        GTE-Qwen2-7B-Instruct (7B) & 33.40 & 39.40 & 34.19 & 42.31 & 63.14 & 70.44 & 42.50 & 51.79 &  57.77 & 65.21 \\  
        Qwen3-Embedding-8B & 39.50 & 46.47 & 36.75 & 46.15 & 60.95 & 71.53 & 44.46 & 55.00 & 60.18 & 65.21 \\       
        \textsc{SweRankEmbed}-Large (7B) & 41.92 & 49.18 & 39.74 & 50.85 & 71.90 & 82.12 & 55.18 & 63.21 &  65.65 & 74.18\\
        \textsc{SweRankEmbedPython}-Large (8B) & 44.24 & 51.98 & 46.15 & 55.98 & 73.72 & 83.94 & 55.71 & 65.00 & 66.96 & 75.05 \\
        \textbf{\textsc{SweRankEmbedMulti}-Large} (8B) & \textbf{46.56} & \textbf{53.73} & \textbf{50.43} & \textbf{62.39} & \textbf{77.37} & \textbf{86.86} & \textbf{56.43} & \textbf{65.36} & \textbf{68.71} & \textbf{76.37} \\
        \bottomrule
    \end{tabular}
    }
    \caption{Table comparing function localization performance of different retrievers in comparable sizes across benchmarks with various programming languages. Our \textsc{SweRankEmbedMulti} models considerably improve performance on both multilingual and python-specific benchmarks.}
    \label{tab:retriever_numbers}
\end{table*}

\begin{table*}[t]
    \centering
    \renewcommand{\arraystretch}{1.2}
    \small
    \resizebox{1\textwidth}{!}{
    \begin{tabular}{l cc cc cc cc cc}
        \toprule
        
        \multirow{3}{*}{\textbf{Model}} &  \multicolumn{4}{c}{\textbf{Multilingual}} & \multicolumn{6}{c}{\textbf{Python}} \\
        \cmidrule(lr){2-5} \cmidrule(lr){6-11}
        
        &\multicolumn{2}{c}{\textbf{SWE-PolyBench}}  & \multicolumn{2}{c}{\textbf{SWE-Bench-Multilingual}} &\multicolumn{2}{c}{\textbf{SWE-Bench-Lite}} & \multicolumn{2}{c}{\textbf{LocBench}} & \multicolumn{2}{c}{\textbf{SWE-Bench-Verified}}\\
        \cmidrule(lr){2-3} \cmidrule(lr){4-5} \cmidrule(lr){6-7} \cmidrule(lr){8-9} \cmidrule(lr){10-11} 
         & \textbf{Acc@5} & \textbf{Acc@10} & \textbf{Acc@5} & \textbf{Acc@10} & \textbf{Acc@5} & \textbf{Acc@10}& \textbf{Acc@5} & \textbf{Acc@10}& \textbf{Acc@5} & \textbf{Acc@10} \\
        \midrule
         SweRankEmbedMulti-Small (0.6B) & 39.01 & 47.14 & 43.16 & 52.56 & 66.79 & 76.28 & 51.25 & 58.93 & 64.99 & 71.77 \\
         + CodeRankLLM (7B) & 42.40 & 51.11 & 48.72 & 58.55 & 72.99 & 80.29 & 55.54 & 63.93 & 66.52 & 75.93 \\
         + Qwen3-Instruct-8B & 43.66 & 51.98 & 50.85 & 59.83 & 72.27 & 80.29 & 56.96 & 64.46 & 68.49 & 75.27\\
         + SweRankLLM-Small (7B) & 51.21 & 56.92 & 55.13 & 63.68 & 77.37 & 85.04 & \textbf{63.39} & \textbf{69.64} & \textbf{73.30} & 77.68 \\
         + \textbf{SweRankLLMMulti-Small (7B)} & \textbf{53.15} & \textbf{59.54} & \textbf{56.41} & \textbf{66.67} & \textbf{80.29} & \textbf{85.77} & 63.04 & 69.46 & \textbf{73.30} & \textbf{79.43} \\
         \midrule
        SweRankEmbedMulti-Large (8B) & 46.56 & 53.73 & 50.43 & 62.39 & 77.37 & 86.86 & 56.43 & 65.36 & 68.71 & 76.37 \\
        + GPT-4.1 & 55.66 & 61.76 & 62.39 & 70.94 & 79.93 & 88.69 & 65.89 & 70.89 & 76.81 & \textbf{81.62} \\
        + SweRankLLM-Large (32B) & 56.73 & 62.73 & 57.69 & 70.51 & 83.58  & 89.42 & 64.64  &  71.25 & 75.49 & 80.96  \\
        + \textbf{SweRankLLMMulti-Large (32B)} & \textbf{58.28} & \textbf{63.21} & \textbf{64.10} & \textbf{71.37} & \textbf{85.40} & \textbf{89.78} & \textbf{66.43} & \textbf{71.96} & \textbf{78.12} & 81.18\\
        \bottomrule
    \end{tabular}
    }
    \caption{Table comparing function localization performance of different rerankers across benchmarks with various programming languages.}
    \label{tab:reranker_numbers}
\end{table*}

\subsection{Datasets \& Metrics}

Our multilingual evaluation data comprises three datasets: SWE-PolyBench~\cite{Rashid2025SWEPolyBenchAM}, SWE-Bench-Multilingual~\cite{yang2025swesmith}, and Multi-SWE-Bench~\cite{Zan2025MultiSWEbenchAM}.  Following~\cite{suresh2024cornstack} and ~\cite{reddy2025swerank}, we transform each (PR, codebase) instance pair from these datasets into the localization format. The PR's corresponding github issue description serves as the retrieval query. The Tree-sitter parsing tool is employed to extract all functions from the codebase, creating candidate corpus. Functions modified within the PR are labeled are considered as positives.
 Moreover, we also consider python-specific benchmarks, specifically SWE-Bench-Lite~\cite{jimenez2023swe}, LocBench~\cite{chen2025locagent} and SWE-Bench-Verified~\cite{chowdhury2024swebenchverified}. 
 
We employ Accuracy at k (Acc@k) for evaluation. This metric deems localization successful if all relevant code locations are correctly identified within the top-k results.

\subsection{SweRankMulti}
\label{sec:swerankmulti_expts}

\subsubsection{Setup}

\paragraph{Model Training:} We train the \textsc{SweRankEmbed} model in two sizes: \textit{small} and \textit{large}. Both models are trained on the \textsc{SweLocMulti} dataset (\S{\ref{sec:swelocmulti}}), with small and large variants initialized with the 0.6B and 8B variants of Qwen3-Embedding~\citep{zhang2025qwen3} respectively. Following \textsc{SweRank}~\cite{reddy2025swerank}, the \textsc{SweRankLLMMulti} small and large rerankers are initialized with CodeRankLLM (7B)~\cite{suresh2024cornstack} and Qwen-2.5-32B-Instruct~\cite{Yang2024Qwen25TR} respectively, and finetuned with \textsc{SweLocMulti} for listwise reranking.

\paragraph{\textsc{SweRankEmbedMulti} Baselines:} For the retriever evaluation, we compare against existing code embedding models such \textsc{SweRankEmbed}~\citep{reddy2025swerank}, and \textsc{CodeRankEmbed}~\citep{suresh2024cornstack}. Since \textsc{SweRankEmbed} was finetuned on GTE-Qwen2-7B-Instruct, we finetune the Qwen3-Embedding models on the python-specific \textsc{SweLoc} dataset~\citep{reddy2025swerank} to get the \textsc{SweRankEmbedPython} variants for better comparison.  We also include Gemini-Embedding~\citep{lee2025gemini}, the current top model on the MTEB leaderboard, as a general-purpose closed-source baseline. 

\paragraph{\textsc{SweRankLLMMulti} Baselines:} For the reranker evaluation, we compare against listwise reranker models such as \textsc{SweRankLLM}~\cite{reddy2025swerank}, \textsc{CodeRankLLM}~\cite{suresh2024cornstack}, and the zero-shot Qwen3-Instruct-8B model~\cite{yang2025qwen3} along with GPT-4.1. 

\begin{figure*}
    \centering
    \includegraphics[width=1.0\linewidth]{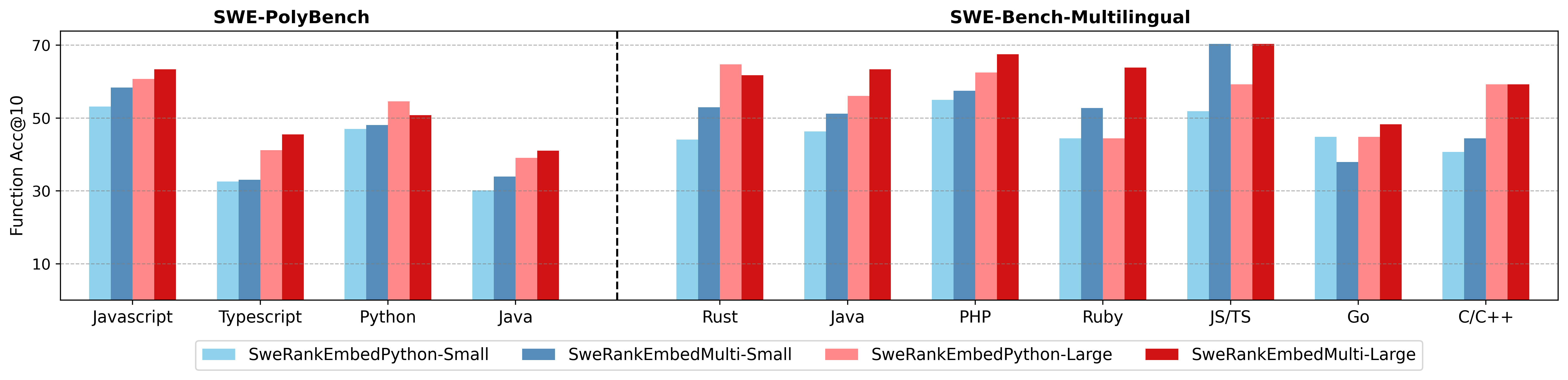}
    \caption{Language-wise function localization accuracy for SWE-PolyBench and SWE-Bench-Multilingual.}
    \label{fig:retriever_language_wise}
\end{figure*}

\begin{table*}[t]
    \centering
    \renewcommand{\arraystretch}{1.2}
    \small
    \resizebox{1\textwidth}{!}{
    \begin{tabular}{l c c c c c}
        \toprule
        
        \multirow{2}{*}{\textbf{Approach}} &  \multicolumn{2}{c}{\textbf{Multilingual}} & \multicolumn{3}{c}{\textbf{Python}}\\
        \cmidrule(lr){2-3} \cmidrule(lr){4-6}
        
        &\multicolumn{1}{c}{\textbf{SWE-PolyBench}}  & \multicolumn{1}{c}{\textbf{SWE-Bench-Multilingual}} & \multicolumn{1}{c}{\textbf{SWE-Bench-Lite}} & \multicolumn{1}{c}{\textbf{LocBench}} & \multicolumn{1}{c}{\textbf{SWE-Bench-Verified}}\\
        \midrule
        
        \textsc{SweRankMulti} (Single-Query) & 59.54 & 66.67 &  85.77 &  69.46 & 79.43\\
        Reformulate (Single-Turn, Multi-Query) & 58.08~\scriptsize(\textcolor{red}{-1.46}) & 64.96~\scriptsize(\textcolor{red}{-1.71}) & 85.77 & 69.86~\scriptsize(\textcolor{mgreen}{+0.40}) & 78.99~\scriptsize(\textcolor{red}{-0.44})\\
        \textsc{SweRankAgent} (Multi-Turn) & \textbf{62.63}~\scriptsize(\textcolor{mgreen}{+3.09}) & \textbf{73.08}~\scriptsize(\textcolor{mgreen}{+6.41}) & \textbf{88.32}~\scriptsize(\textcolor{mgreen}{+2.55}) & \textbf{70.91}~\scriptsize(\textcolor{mgreen}{+1.45}) & \textbf{81.74}~\scriptsize(\textcolor{mgreen}{+2.31}) \\
        \bottomrule
    \end{tabular}
    }
    \caption{Table comparing \textsc{SweRankAgent}'s multi-turn function localization performance (Acc@10) against the single-turn \textsc{SweRankMulti} tool in addition to a multi-query reformulation approach.}
    \label{tab:swerankagent_numbers}
\end{table*}

\subsubsection{Results}

Table~\ref{tab:retriever_numbers} shows the function localization performance different embedding models when compared at similar size ranges. We see that \textsc{SweRankEmbedMulti} achieves SOTA performance at both size variants, considerably improving upon the existing \textsc{SweRankEmbed}. Moreover, we see that a better base embedding model (\textsc{SweRankEmbed} vs \textsc{SweRankEmbedPython}) and training with multilingual data (\textsc{SweRankEmbedPython} vs \textsc{SweRankEmbedMulti}) both contribute to improvements in performance. Interestingly, \textsc{SweRankEmbedMulti} even improves over \textsc{SweRankEmbedPython} on python-specific benchmarks, despite the latter being trained on the python-exclusive \textsc{SweLoc}. This could be attributed to better generalization performance from including more languages in the training data. 

Table~\ref{tab:reranker_numbers} shows the function localization performance of the reranker when used with retrievers of different sizes. We see that \textsc{SweRankLLMMulti} reranker consistently improve localization performance over the \textsc{SweRankLLM} models, with the large variant even outperforming GPT-4.1. 

\paragraph{Language-wise Performance:} Figure~\ref{fig:retriever_language_wise} shows the function localization performance of the retrievers separately for each of the languages in SWE-PolyBench and SWE-Bench-Multilingual. We observe that both the small and large variants of \textsc{SweRankEmbed} consistently improve performance on most languages, compared to \textsc{SweRankEmbedPython}. 

\subsection{SweRankAgent}
\label{sec:swerankagent_expts}

\subsubsection{Setup}

\paragraph{\textsc{SweRankMulti} tool:} We equip the \textsc{SweRankAgent} framework with a search tool that, when given with a query in the form of an issue description, returns the top-10 localized functions within the codebase. The \textsc{SweRankMultiEmbed} retriever first obtains the top-100 results which are then passed to the \textsc{SweRankMultiLLM} reranker. For efficiency considerations, the codebase function embeddings for the retriever are pre-cached. Given compute considerations, we only use the small variants of the retriever and reranker for the \textsc{SweRankMulti} tool. 

\paragraph{Baselines:} To investigate the performance of a multi-query setup, we design a \textit{Reformulate} baseline that does query-reformulation to generate multiple queries from the original issue description. In total, we use the original issue description as the query plus five reformulated queries generated by GPT-5. For each query, results from the \textsc{SweRankMulti} tool are collected, and the aggregated output is then again passed to the \textsc{SweRankMultiLLM} reranker to obtain final top-10 function localization result. Moreover, we also use the \textsc{SweRankMulti} tool as a single-query baseline.

\begin{figure*}
    \centering
    \begin{subfigure}[b]{0.49\textwidth}
        \centering
        \includegraphics[width=1.0\linewidth]{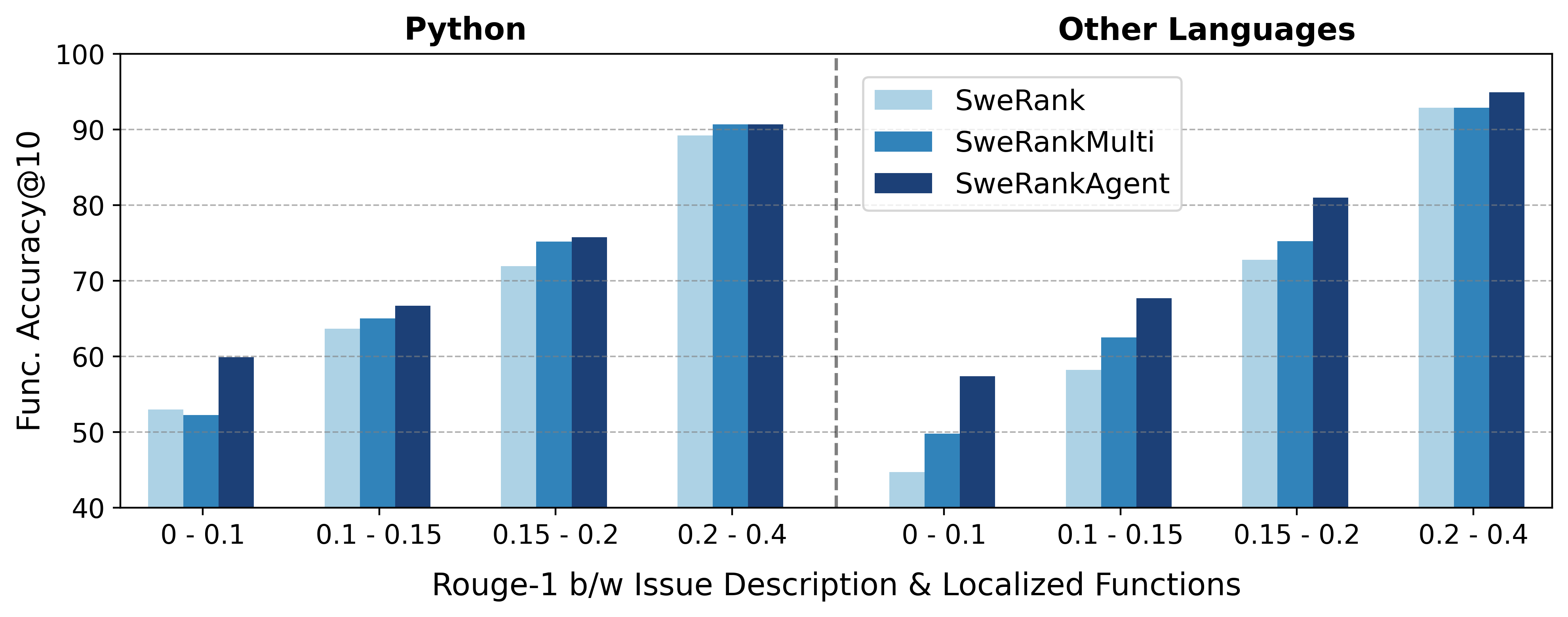}
        \caption{Performance breakdown by lexical overlap.}
        \label{fig:sub1}
    \end{subfigure}
    \hfill 
    \begin{subfigure}[b]{0.49\textwidth}
        \centering
        \includegraphics[width=1.0\linewidth]{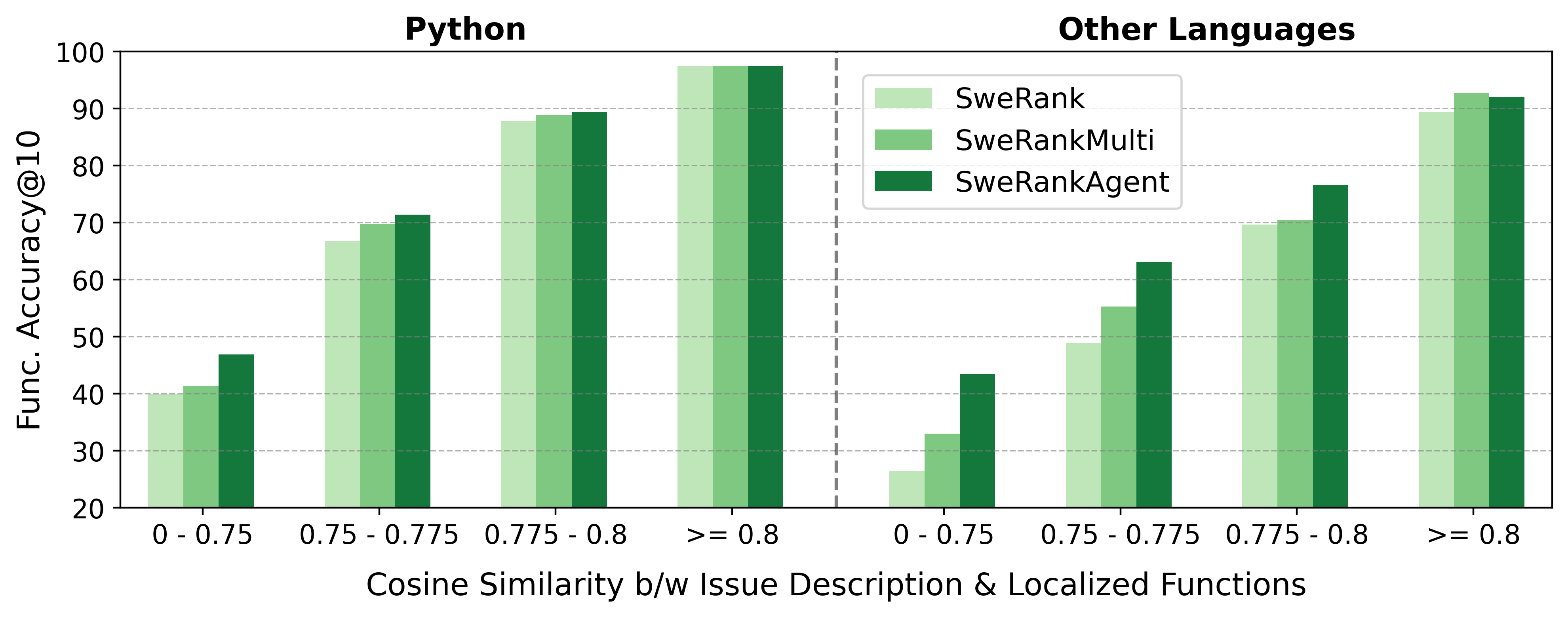}
        \caption{Performance breakdown by semantic overlap.}
        \label{fig:sub2}
    \end{subfigure}
    \caption{Comparison of \textsc{SweRank}, \textsc{SweRankMulti}, and \textsc{SweRankAgent} performance across varying levels of difficulty. The plots show Function Accuracy@10 separately for Python and other languages, broken down by (a) lexical overlap using ROUGE-1 scores and (b) semantic overlap using cosine similarity.}
    \label{fig:overlap}
\end{figure*}

\subsubsection{Results}

Table~\ref{tab:swerankagent_numbers} shows the function localization performance of \textsc{SweRankAgent} using GPT-5 as the underlying LLM. 
Firstly, we observe that the multi-query \textit{Reformulate} baseline yields lower performance than the single-query baseline. This suggests that directly using reformulated queries introduces additional noise, highlighting the inherent challenges in improving coverage of localization candidates with multiple queries.

In contrast, \textsc{SweRankAgent} consistently improves localization performance across all datasets, with more than a 3-point boost on SWE-PolyBench and over a 6-point boost on SWE-Bench-Multilingual. 





\paragraph{Qualitative Example:} Figure~\ref{fig:swerank_agent} illustrates the advantage of multi-turn issue localization through a qualitative example where iterative search succeeds while single-pass retrieval fails.

At first glance, the issue appears straightforward: the pylint \textit{re.error} traceback points to \textit{re.compile}. A naive single-pass retrieval would likely stop at \textit{$\_$regexp$\_$validator}, the immediate crash site but not the root cause. The real bug lies upstream, in how a comma-separated list of regular expressions is parsed before validation. Modifying \textit{$\_$regexp$\_$validator} would thus be incorrect; the fix belongs in the function handling the comma-separated-value (CSV) input. \textsc{SweRankAgent}'s multi-turn trajectory below demonstrates how iterative refinement enables correct localization.

\begin{itemize}[leftmargin=*]
\item \textbf{Turn 1: Broad Exploration.}
The agent begins with a broad query guided by the traceback to locate where configuration options are defined and validated.
This identifies relevant configuration modules (\textit{arguments$\_$manager.py}, \textit{utils.py}) and the symptom function \textit{$\_$regexp$\_$validator}, but not the root cause. A single-shot system would likely stop here.

\item \textbf{Turn 2: Contextual Refinement.}  
After examining the initial results, the agent infers that the issue relates to how naming options are validated.  
It refines its search to focus on functions involved in naming configuration and regular expression validation, uncovering key functions such as \textit{$\_$create$\_$naming$\_$options} and \textit{$\_$regexp$\_$csv$\_$validator}, revealing the CSV-handling pathway.

\item \textbf{Turn 3: Pinpointing the Target.}  
The agent then hypothesizes the existence of a transformer function that processes CSV inputs prior to validation. A focused search on transformer–validator interactions finally retrieves the correct target:\textit{$\_$regexp$\_$csv$\_$transformer}.  
\end{itemize}
\subsubsection{Analysis by Issue Complexity}

Aggregate performance metrics can mask how localization methods behave under varying levels of problem difficulty. To better understand where \textsc{SweRank+} provides gains, we analyze performance across two orthogonal dimensions of complexity: (1) the degree of lexical and semantic overlap between the issue description and the target code, and (2) the number of ground-truth localized functions. We primarily aim to investigate whether the agentic search process is particularly beneficial for hard cases with multiple target functions or requiring reasoning beyond surface-level textual similarity.

\paragraph{Performance by Overlap:} We bucket instances based on lexical overlap (ROUGE-1) and semantic overlap (cosine similarity) between the issue description and the ground-truth localized functions, with results shown in Figure~\ref{fig:overlap}.  We can see that performance generally improves with higher overlap, indicating that keyword- or semantically-aligned issues are easier to localize. Nonetheless, \textsc{SweRankAgent} demonstrates clear advantages in low-overlap settings. In buckets with minimal lexical or semantic overlap, where naive keyword matching is insufficient, the agent consistently shows bigger improvements compared to buckets with more overlap. The agent's ability to reformulate queries and reason over intermediate results allows it to bridge the gap when the issue description lacks specific keywords present in the target functions, demonstrating that the multi-turn reasoning is  crucial for traversing the ``semantic gap'' in harder instances where the failure description is semantically distant from the root cause.

\begin{figure}[t]
    \centering
    \includegraphics[width=1.0\linewidth]{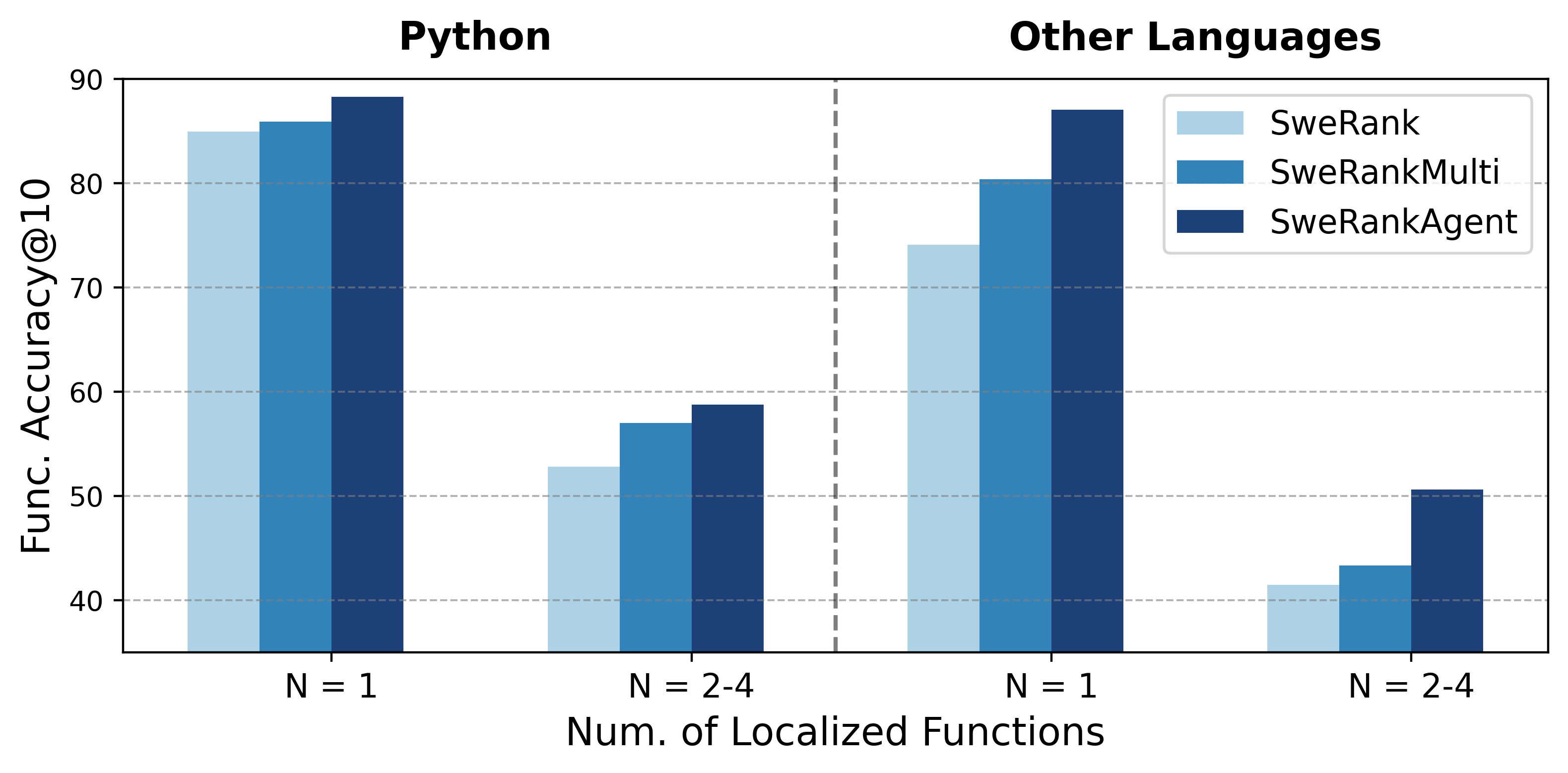}
    \caption{Function Accuracy@10 breakdown by number of target localization functions.}
    \label{fig:issue_complexity}
\end{figure}

\paragraph{Performance by \# of Target Functions:} We stratify test instances by the number of functions modified in the ground-truth patch, which serves as a proxy for issue complexity. As seen in Figure~\ref{fig:issue_complexity}, localization accuracy degrades as the number of target localization functions increases, reflecting the increased difficulty of multi-function bugs and feature requests. While the multilingual training of \textsc{SweRankMulti} does considerably improve performance for other languages, \textsc{SweRankMulti} particularly shines in more complex multi-function localization settings.

\section{Conclusion}

In this work, we introduced \textsc{SweRank+}, a comprehensive software issue localization framework that overcomes the limitations of Python-centric and single-pass ranking approaches. Our contributions are twofold. First, we developed \textsc{SweRankMulti}, a multilingual retrieve-and-rerank system trained on the newly curated \textsc{SweLocMulti} dataset, that establishes new state-of-the-art performance on multilingual benchmarks while maintaining strong efficacy on Python-specific tasks. Second, we proposed \textsc{SweRankAgent}, a lightweight agentic framework that employs iterative, multi-turn reasoning to refine localization candidates, thereby consistently outperforming single-pass ranking approaches. Our analysis highlights that this agentic approach is particularly effective for complex issues characterized by low overlap between issue description and target functions or those requiring modifications across multiple functions. This work emphasizes a hybrid design that bridges the gap between efficient retrieval systems and computationally heavy agentic frameworks, demonstrating that lightweight, iterative reasoning can effectively solve complex localization problems. Future work will explore integrating \textsc{SweRank+} into end-to-end automated program repair pipelines.


\bibliography{custom}

\begin{thebibliography}{29}
\providecommand{\natexlab}[1]{#1}

\bibitem[{Chang et~al.(2025)Chang, Zhou, Wang, Lo, and Li}]{chang2025bridging}
Jianming Chang, Xin Zhou, Lulu Wang, David Lo, and Bixin Li. 2025.
\newblock Bridging bug localization and issue fixing: A hierarchical localization framework leveraging large language models.
\newblock \emph{arXiv preprint arXiv:2502.15292}.

\bibitem[{Chen et~al.(2025)Chen, Tang, Deng, Wu, Wu, Jiang, Prasanna, Cohan, and Wang}]{chen2025locagent}
Zhaoling Chen, Xiangru Tang, Gangda Deng, Fang Wu, Jialong Wu, Zhiwei Jiang, Viktor Prasanna, Arman Cohan, and Xingyao Wang. 2025.
\newblock Locagent: Graph-guided llm agents for code localization.
\newblock \emph{arXiv preprint arXiv:2503.09089}.

\bibitem[{Chowdhury et~al.(2024)Chowdhury, Aung, Shern, Jaffe, Sherburn, Starace, Mays, Dias, Aljubeh, Glaese, Jimenez, Yang, Ho, Patwardhan, Liu, and Madry}]{chowdhury2024swebenchverified}
Neil Chowdhury, James Aung, Chan~Jun Shern, Oliver Jaffe, Dane Sherburn, Giulio Starace, Evan Mays, Rachel Dias, Marwan Aljubeh, Mia Glaese, Carlos~E. Jimenez, John Yang, Leyton Ho, Tejal Patwardhan, Kevin Liu, and Aleksander Madry. 2024.
\newblock \href {https://openai.com/index/introducing-swe-bench-verified/} {Introducing {SWE}-bench verified}.

\bibitem[{Dong et~al.(2025)Dong, Jiang, Qian, Wang, Zhang, Jin, and Li}]{dong2025surveycodegenerationllmbased}
Yihong Dong, Xue Jiang, Jiaru Qian, Tian Wang, Kechi Zhang, Zhi Jin, and Ge~Li. 2025.
\newblock \href {https://arxiv.org/abs/2508.00083} {A survey on code generation with llm-based agents}.
\newblock \emph{Preprint}, arXiv:2508.00083.

\bibitem[{He et~al.(2025)He, Treude, and Lo}]{he2025llm}
Junda He, Christoph Treude, and David Lo. 2025.
\newblock Llm-based multi-agent systems for software engineering: Literature review, vision, and the road ahead.
\newblock \emph{ACM Transactions on Software Engineering and Methodology}, 34(5):1--30.

\bibitem[{Jimenez et~al.(2024)Jimenez, Yang, Wettig, Yao, Pei, Press, and Narasimhan}]{jimenez2023swe}
Carlos~E Jimenez, John Yang, Alexander Wettig, Shunyu Yao, Kexin Pei, Ofir Press, and Karthik~R Narasimhan. 2024.
\newblock \href {https://openreview.net/forum?id=VTF8yNQM66} {{SWE}-bench: Can language models resolve real-world github issues?}
\newblock In \emph{The Twelfth International Conference on Learning Representations}.

\bibitem[{Jones and Harrold(2005)}]{jones2005empirical}
James~A Jones and Mary~Jean Harrold. 2005.
\newblock Empirical evaluation of the tarantula automatic fault-localization technique.
\newblock In \emph{Proceedings of the 20th IEEE/ACM international Conference on Automated software engineering}, pages 273--282.

\bibitem[{Lee et~al.(2025)Lee, Chen, Dua, Cer, Shanbhogue, Naim, {\'A}brego, Li, Chen, Vera et~al.}]{lee2025gemini}
Jinhyuk Lee, Feiyang Chen, Sahil Dua, Daniel Cer, Madhuri Shanbhogue, Iftekhar Naim, Gustavo~Hern{\'a}ndez {\'A}brego, Zhe Li, Kaifeng Chen, Henrique~Schechter Vera, and 1 others. 2025.
\newblock Gemini embedding: Generalizable embeddings from gemini.
\newblock \emph{arXiv preprint arXiv:2503.07891}.

\bibitem[{Li et~al.(2019)Li, Li, Zhang, and Zhang}]{li2019deepfl}
Xia Li, Wei Li, Yuqun Zhang, and Lingming Zhang. 2019.
\newblock Deepfl: Integrating multiple fault diagnosis dimensions for deep fault localization.
\newblock In \emph{Proceedings of the 28th ACM SIGSOFT international symposium on software testing and analysis}, pages 169--180.

\bibitem[{Oord et~al.(2018)Oord, Li, and Vinyals}]{oord2018representation}
Aaron van~den Oord, Yazhe Li, and Oriol Vinyals. 2018.
\newblock Representation learning with contrastive predictive coding.
\newblock In \emph{Advances in Neural Information Processing Systems}, pages 10203--10213.

\bibitem[{Qin et~al.(2024)Qin, Jagerman, Hui, Zhuang, Wu, Yan, Shen, Liu, Liu, Metzler et~al.}]{qin2024large}
Zhen Qin, Rolf Jagerman, Kai Hui, Honglei Zhuang, Junru Wu, Le~Yan, Jiaming Shen, Tianqi Liu, Jialu Liu, Donald Metzler, and 1 others. 2024.
\newblock Large language models are effective text rankers with pairwise ranking prompting.
\newblock In \emph{Findings of the Association for Computational Linguistics: NAACL 2024}, pages 1504--1518.

\bibitem[{Rashid et~al.(2025)Rashid, Bock, Yuan, Buccholz, Esler, Valentin, Franceschi, Wistuba, Sivaprasad, Kim, Deoras, Zappella, and Callot}]{Rashid2025SWEPolyBenchAM}
Muhammad~Shihab Rashid, Christian Bock, Zhuang Yuan, Alexander Buccholz, Tim Esler, Simon Valentin, Luca Franceschi, Martin Wistuba, Prabhu~Teja Sivaprasad, Woo~Jung Kim, Anoop Deoras, Giovanni Zappella, and Laurent Callot. 2025.
\newblock \href {https://api.semanticscholar.org/CorpusID:277741387} {Swe-polybench: A multi-language benchmark for repository level evaluation of coding agents}.
\newblock \emph{ArXiv}, abs/2504.08703.

\bibitem[{Reddy et~al.(2025)Reddy, Suresh, Doo, Liu, Nguyen, Zhou, Yavuz, Xiong, Ji, and Joty}]{reddy2025swerank}
Revanth~Gangi Reddy, Tarun Suresh, JaeHyeok Doo, Ye~Liu, Xuan~Phi Nguyen, Yingbo Zhou, Semih Yavuz, Caiming Xiong, Heng Ji, and Shafiq Joty. 2025.
\newblock Swerank: Software issue localization with code ranking.
\newblock \emph{arXiv preprint arXiv:2505.07849}.

\bibitem[{Suresh et~al.(2024)Suresh, Reddy, Xu, Nussbaum, Mulyar, Duderstadt, and Ji}]{suresh2024cornstack}
Tarun Suresh, Revanth~Gangi Reddy, Yifei Xu, Zach Nussbaum, Andriy Mulyar, Brandon Duderstadt, and Heng Ji. 2024.
\newblock Cornstack: High-quality contrastive data for better code ranking.
\newblock \emph{arXiv preprint arXiv:2412.01007}.

\bibitem[{Team et~al.(2024)Team, Zhao, Hui, Howland, Nguyen, Zuo, Hu, Choquette-Choo, Shen, Kelley, Bansal, Vilnis, Wirth, Michel, Choy, Joshi, Kumar, Hashmi, Agrawal, Gong, Fine, Warkentin, Hartman, Ni, Korevec, Schaefer, and Huffman}]{codegemmateam2024codegemmaopencodemodels}
CodeGemma Team, Heri Zhao, Jeffrey Hui, Joshua Howland, Nam Nguyen, Siqi Zuo, Andrea Hu, Christopher~A. Choquette-Choo, Jingyue Shen, Joe Kelley, Kshitij Bansal, Luke Vilnis, Mateo Wirth, Paul Michel, Peter Choy, Pratik Joshi, Ravin Kumar, Sarmad Hashmi, Shubham Agrawal, and 8 others. 2024.
\newblock \href {https://arxiv.org/abs/2406.11409} {Codegemma: Open code models based on gemma}.
\newblock \emph{Preprint}, arXiv:2406.11409.

\bibitem[{Torun et~al.(2025)Torun, Demircan, G{\"o}n, and T{\"u}z{\"u}n}]{torun2025past}
Utku~Boran Torun, Mehmet~Taha Demircan, Mahmut~Furkan G{\"o}n, and Eray T{\"u}z{\"u}n. 2025.
\newblock Past, present, and future of bug tracking in the generative ai era.
\newblock \emph{arXiv preprint arXiv:2510.08005}.

\bibitem[{Wong et~al.(2016)Wong, Gao, Li, Abreu, and Wotawa}]{wong2016survey}
W~Eric Wong, Ruizhi Gao, Yihao Li, Rui Abreu, and Franz Wotawa. 2016.
\newblock A survey on software fault localization.
\newblock \emph{IEEE Transactions on Software Engineering}, 42(8):707--740.

\bibitem[{Yang et~al.(2025{\natexlab{a}})Yang, Li, Yang, Zhang, Hui, Zheng, Yu, Gao, Huang, Lv et~al.}]{yang2025qwen3}
An~Yang, Anfeng Li, Baosong Yang, Beichen Zhang, Binyuan Hui, Bo~Zheng, Bowen Yu, Chang Gao, Chengen Huang, Chenxu Lv, and 1 others. 2025{\natexlab{a}}.
\newblock Qwen3 technical report.
\newblock \emph{arXiv preprint arXiv:2505.09388}.

\bibitem[{Yang et~al.(2024{\natexlab{a}})Yang, Jimenez, Wettig, Lieret, Yao, Narasimhan, and Press}]{yang2024sweagent}
John Yang, Carlos~E Jimenez, Alexander Wettig, Kilian Lieret, Shunyu Yao, Karthik Narasimhan, and Ofir Press. 2024{\natexlab{a}}.
\newblock Swe-agent: Agent-computer interfaces enable automated software engineering.
\newblock \emph{arXiv preprint arXiv:2405.15793}.

\bibitem[{Yang et~al.(2025{\natexlab{b}})Yang, Lieret, Jimenez, Wettig, Khandpur, Zhang, Hui, Press, Schmidt, and Yang}]{yang2025swesmith}
John Yang, Kilian Lieret, Carlos~E. Jimenez, Alexander Wettig, Kabir Khandpur, Yanzhe Zhang, Binyuan Hui, Ofir Press, Ludwig Schmidt, and Diyi Yang. 2025{\natexlab{b}}.
\newblock \href {https://arxiv.org/abs/2504.21798} {Swe-smith: Scaling data for software engineering agents}.
\newblock \emph{Preprint}, arXiv:2504.21798.

\bibitem[{Yang et~al.(2024{\natexlab{b}})Yang, Yang, Zhang, Hui, Zheng, Yu, Li, Liu, Huang, Dong, Wei, Lin, Yang, Tu, Zhang, Yang, Yang, Zhou, Lin, Dang, Lu, Bao, Yang, Yu, Li, Xue, Zhang, Zhu, Men, Lin, Li, Xia, Ren, Ren, Fan, Su, Zhang, Wan, Liu, Cui, Zhang, Qiu, Quan, and Wang}]{Yang2024Qwen25TR}
Qwen~An Yang, Baosong Yang, Beichen Zhang, Binyuan Hui, Bo~Zheng, Bowen Yu, Chengyuan Li, Dayiheng Liu, Fei Huang, Guanting Dong, Haoran Wei, Huan Lin, Jian Yang, Jianhong Tu, Jianwei Zhang, Jianxin Yang, Jiaxin Yang, Jingren Zhou, Junyang Lin, and 25 others. 2024{\natexlab{b}}.
\newblock \href {https://api.semanticscholar.org/CorpusID:274859421} {Qwen2.5 technical report}.
\newblock \emph{ArXiv}, abs/2412.15115.

\bibitem[{Yao et~al.(2023)Yao, Zhao, Yu, Du, Shafran, Narasimhan, and Cao}]{yao2023react}
Shunyu Yao, Jeffrey Zhao, Dian Yu, Nan Du, Izhak Shafran, Karthik Narasimhan, and Yuan Cao. 2023.
\newblock React: Synergizing reasoning and acting in language models.
\newblock In \emph{International Conference on Learning Representations (ICLR)}.

\bibitem[{Yaraghi et~al.(2025)Yaraghi, Gharachorlu, Fatima, Briand, Wan, and Gao}]{yaraghi2025black}
Ahmadreza~Saboor Yaraghi, Golnaz Gharachorlu, Sakina Fatima, Lionel~C Briand, Ruiyuan Wan, and Ruifeng Gao. 2025.
\newblock Black-box test code fault localization driven by large language models and execution estimation.
\newblock \emph{arXiv preprint arXiv:2506.19045}.

\bibitem[{Yu et~al.(2025)Yu, Zhang, Zhao, Huang, Yao, Ding, and Zhao}]{yu2025orcaloca}
Zhongming Yu, Hejia Zhang, Yujie Zhao, Hanxian Huang, Matrix Yao, Ke~Ding, and Jishen Zhao. 2025.
\newblock Orcaloca: An llm agent framework for software issue localization.
\newblock \emph{arXiv preprint arXiv:2502.00350}.

\bibitem[{Zan et~al.(2025)Zan, Huang, Liu, Chen, Zhang, Xin, Chen, Liu, Zhong, Li, Liu, Xiao, Chen, Zhang, Su, Liu, Long, Shen, and Xiang}]{Zan2025MultiSWEbenchAM}
Daoguang Zan, Zhirong Huang, Wei Liu, Hanwu Chen, Linhao Zhang, Shulin Xin, Lu~Chen, Qi~Liu, Xiaojian Zhong, Aoyan Li, Siyao Liu, Yongsheng Xiao, Liangqiang Chen, Yuyu Zhang, Jing Su, Tianyu Liu, Rui Long, Kai Shen, and Liang Xiang. 2025.
\newblock \href {https://api.semanticscholar.org/CorpusID:277510042} {Multi-swe-bench: A multilingual benchmark for issue resolving}.
\newblock \emph{ArXiv}, abs/2504.02605.

\bibitem[{Zeller(2002)}]{zeller2002isolating}
Andreas Zeller. 2002.
\newblock Isolating cause-effect chains from computer programs.
\newblock \emph{ACM SIGSOFT Software Engineering Notes}, 27(6):1--10.

\bibitem[{Zhang et~al.(2025)Zhang, Li, Long, Zhang, Lin, Yang, Xie, Yang, Liu, Lin et~al.}]{zhang2025qwen3}
Yanzhao Zhang, Mingxin Li, Dingkun Long, Xin Zhang, Huan Lin, Baosong Yang, Pengjun Xie, An~Yang, Dayiheng Liu, Junyang Lin, and 1 others. 2025.
\newblock Qwen3 embedding: Advancing text embedding and reranking through foundation models.
\newblock \emph{arXiv preprint arXiv:2506.05176}.

\bibitem[{Zhuang et~al.(2024)Zhuang, Qin, Hui, Wu, Yan, Wang, and Bendersky}]{zhuang2024beyond}
Honglei Zhuang, Zhen Qin, Kai Hui, Junru Wu, Le~Yan, Xuanhui Wang, and Michael Bendersky. 2024.
\newblock Beyond yes and no: Improving zero-shot llm rankers via scoring fine-grained relevance labels.
\newblock In \emph{Proceedings of the 2024 Conference of the North American Chapter of the Association for Computational Linguistics: Human Language Technologies (Volume 2: Short Papers)}, pages 358--370.

\bibitem[{Zhuang et~al.(2023)Zhuang, Qin, Jagerman, Hui, Ma, Lu, Ni, Wang, and Bendersky}]{zhuang2023rankt5}
Honglei Zhuang, Zhen Qin, Rolf Jagerman, Kai Hui, Ji~Ma, Jing Lu, Jianmo Ni, Xuanhui Wang, and Michael Bendersky. 2023.
\newblock Rankt5: Fine-tuning t5 for text ranking with ranking losses.
\newblock In \emph{Proceedings of the 46th International ACM SIGIR Conference on Research and Development in Information Retrieval}, pages 2308--2313.

\end{thebibliography}

\appendix

\begin{tcolorbox}[
    title=\textsc{SweRankLLMMulti} Prompt,                 
    colback=blue!5!white,              
    colframe=blue!75!black,            
    fonttitle=\bfseries,               
    sharp corners,                     
    boxrule=1pt,                       
    left=6pt, right=6pt, top=4pt, bottom=4pt, 
    before skip=8pt, after skip=8pt,   
]
\small
\#\# \textbf{System Prompt}\\
\\
You are CodeRanker, an intelligent code reviewer that can analyze GitHub issues and rank code functions based on their relevance to containing the faults causing the GitHub issue.\\
\\
\#\# \textbf{User Prompt}\\
\\
I will provide you with 10 code functions, each indicated by a numerical identifier []. Rank the code functions based on their relevance to containing the faults causing the following GitHub issue:
$<$Issue Description$>$\\
\\
\#\#\# \textbf{Code Functions}\\
\\
$[$1$]$: $<$Function 1$>$\\
$[$2$]$: $<$Function 2$>$\\
... \\
$[$10$]$: $<$Function 10$>$\\
\\
\#\#\# \textbf{Response Format}\\
\\
All the code functions should be included and listed using identifiers, in descending order of relevance. The output format should be [] > [], e.g., [2] > [1]. Only respond with the ranking results, do not give any explanation.
\end{tcolorbox}

\begin{tcolorbox}[
    title=\textsc{SweRankAgent} Prompt,                 
    colback=blue!5!white,              
    colframe=blue!75!black,            
    fonttitle=\bfseries,               
    sharp corners,                     
    boxrule=1pt,                       
    left=6pt, right=6pt, top=4pt, bottom=4pt, 
    before skip=8pt, after skip=8pt,   
]
\small
\#\# \textbf{System Prompt}\\
\\
You are an intelligent assistant specializing in software issue localization. Your primary goal is to identify up to 10 unique code functions from a given codebase that are most likely to require modification to fix a provided software issue. You must operate by iteratively using the search tool at your disposal over a series of steps.\\
\\
\#\#\# \textbf{Rules and Guidelines}\\
\\
1. Understand the Issue carefully.\\
2. Iterative Search: perform sequential `search` calls; number of rounds is configurable.\\
3. Review \& Reflect after each `search`: use results to inform your next query. Avoid duplicates.\\
4. Explain \& Reformulate: explain relevance for new functions, then reason about how to refine the next query.\\
5. Termination: once coverage is sufficient or rounds are done, call `finish` with up to 10 functions.\\
\\
\#\#\# \textbf{Available Tools}\\
\\
"\textbf{name}": "search"\\
"\textbf{description}": "Searches the codebase for functions relevant to the query. Returns a list of candidate functions found based on the description of the issue passed to the tool."\\
"\textbf{parameters}": "issue\_description"\\
\\
"\textbf{name}": "finish"\\
"\textbf{description}": "Call this tool when you are confident you have identified all the top relevant functions."\\
"\textbf{parameters}": null\\
\\
\#\#\# \textbf{Helpful Pointers}\\
\\
1.  Use `search` with complementary angles across rounds.\\
2.  Prefer high-coverage, low-duplicate results.\\
3.  After each `search`, explain why each new function is relevant.\\
4.  Then justify your query reformulation before making the next call.\\
5.  Use `finish` when confident.\\
\\
\#\#\# \textbf{Expected Response Format}\\
\\
Your response MUST follow this format:\\
\\
THOUGHT: Summarize what you just learned from the latest search results. For EACH newly added function, provide a brief relevance explanation describing why it may relate to the issue description.\\
\\
REFORMULATION: Explain how you will adjust the next search query to improve coverage/diversity and reduce duplicates.\\
\\
ACTION:\\
\{"name": "...", "arguments": \{ ... \}\}
\\
\\
\#\# \textbf{User Prompt}\\
\\
$<$ Original Github Issue Description$>$
\end{tcolorbox}

\end{document}